# Exploring the impact of virtual reality user engagement on tourist behavioral response integrated an environment concern of touristic travel perspective: A new hybrid machine learning approach

SHANG D.W.

dshang31@gatech.edu

## Abstract

Due to the impact of the COVID-19 pandemic, new attractions ways are tended to be adapted by compelling sites to provide tours product and services, such as virtual reality (VR) to visitors. Based on a systematic human-computer interaction (HCI) user engagement and Narrative transportation theory, we develop and test a theoretical framework using a hybrid partial least squares structural equation model (PLS-SEM) and artificial neural network (ANN) machine learning approach that examines key user engagement drivers of visitors' imagery and in-person tour intentions (ITI) during COVID-19. Further, we proposed a novel and hybrid approach called Reflective and Formative PLS-SEM-ANN (FRPSA) with considering both reflective and second-order formative constructs in PLS-SEM giving scope to their different advantages in a complex model. According to a sample of visitors' responses, the results demonstrate that a) user engagement, including felt involvement, aesthetic appeal, perceived usability, focused attention, endurability, and novelty, all directly affect in-person tour intentions; b) environment concern of touristic travel (EC) positively moderates the relationships between user engagement and ITI; c) EC negatively moderates the relationships between imagery and ITI; d) imagery exerts the mediating effect between user engagement and ITI; e) the felt involvement and aesthetic appeal show both the linear significance impact and nonlinear importance. Finally, contributions to theories and practical implications are discussed accordingly.

## 1. Introduction

The pandemic has brought unprecedented changes and posed a considerable challenge to tourism (Fan et al., 2022; Lee et al., 2022; Xiong et al., 2021; Zhu et al., 2022). The COVID-19 pandemic has had a profound and lasting impact on 90% of the world's population (Lohr et al., 2022). Given the global pandemic of COVID-19, it is now more urgent than ever for the tourism sector to offer new experiences to attract visitors (Soulard & McGehee, 2022). To ease the pressure, the WORLD Tourism Council recommends boosting the budgets of tourist attractions, especially using virtual reality technology (Ying et al., 2021). During the pandemic, attracting tourists through VR and other technological means has become the content that stakeholders need to explore. Meanwhile, by 2023, the market supported by technology, including VR travel (Statista, 2020), will reach $160 billion. VR can bring consumers a "try before you buy" experience, and after the relaxation of travel restrictions, it will also stimulate tourists' interest in travel (Flavián et al., 2019). In addition, new travel opportunities can be harnessed to address the current limitations of travel behavior as a result of COVID-19, thus providing a new approach for tourists to operate and market. Also, with holidays getting safer, some consumers might want to encourage to get back out. Virtual reality technology allows visitors to experience an intangible travel experience in the virtual world, which will influence customers' visits.

In the past few years, people witnessed the unprecedented acceleration of technology and changes in the digitally virtual world (Beaunoyer & Guitton, 2021; de Regt et al., 2020; Merkx & Nawijn, 2021). VR has been summarised as a technological development that has had a significant impact on the travel industry due to its ability to engage consumers and serve marketing and tourism (Bec et al., 2021). VR is widely used in hotel experiences, destination brands, and cultural traditions (Flavián et al., 2019; Loureiro et al., 2020). In these industries, every application of VR is based on the technology's ability to change people's experiences and positively influence people's behavior (Bec et al., 2021; Zheng et al., 2021). Thus, VR has attracted both practitioners and academia.

Similar to other cutting-edge technologies (i.e., AI), the significance of VR technologies in the tourism field is theoretically confused. Though VR offers opportunities to effectuate destination management, the research on the tourism VR experience is still limited and sometime confused (Loureiro et al., 2020). Although user experience from human-computer interaction (HCI) perspective can lessen the confusion of VR technologies decision-making for stakeholders (O'Brien et al., 2018), less is known about its affecting mechanisms on VR technologies usage, especially for tourist intention in the tourism context. Thus, stakeholders still cannot realize its nature and importance in using it to improve VR technologies diffusion. Specifically, previous studies have made valuable contributions to VR technologies in general and VR tourism in specific literature (Loureiro et al., 2020; O'Brien et al., 2018). However, current research on tourist in-person visit intentions based on VR technologies on antecedents and decision-making process is still inadequate with gaps for the following reasons.

First, different and conflicting views have emerged over VR technologies and visitors' experience and behavioral intention. For example, previous research (Bec et al., 2021; Flavián et al., 2019; Loureiro et al., 2020; Zheng et al., 2021) posits that VR usage is determined by personal characteristics (e.g., user innovativeness and social distancing behavior) or by technology characteristics (e.g., visual design and vividness). A consistent perspective needs to be put forward, and user engagement from human-computer interaction (HCI) perspective provides with understanding and improving the uptake of VR technologies, especially in tourism context during pandemic. Second, although researchers (Alyahya & McLean, 2021; Ying et al., 2021) have offered insights into important perception of technology (e.g., imagery and presence) variables, interaction and moderating effects is neglected or sometime limited.

Moreover, in recent years, the notion of pro-environment has been deeply rooted in people's minds and exerts a subtle influence on individual consumers, including tourists (Champoux-Larsson & Knežević, 2021; Font & McCabe, 2017; Gössling, 2020; MacInnes et al., 2022; Mertena et al., 2022; Suárez-Rojas et al., 2021). Tourism ethics and environment concern play the increasingly significant roles in tourist consumption and response (Malone et al., 2014; Suárez-Rojas et al., 2021; Talwar et al., 2022). For example, previous studies show that environment concern of touristic travel (EC) as a significant indicator in tourism is correlated with tourism intention (Suárez-Rojas et al., 2021; Talwar et al., 2022). To the best of our knowledge, there is no framework has been constructed to understand the moderating effect of environment concern of touristic travel and individual perception (e.g. imagery and HCI user engagement elements) on the individual decision-making process especially in tourism context.

Further, studies have been conducted on individual visitors' response used only linear and compensatory models, less is known about the nonlinear model and expert systems in VR technologies and tourist behavior. Linear model and nonlinear model have their own advantages and disadvantages in research. Only applying either linear model or nonlinear model without assessing hybrid approach, this leaves the gap to examine and bridge in theory building and nonlinear prediction, especially in the context of VR technologies and tourist behavior with the evidence in emerging markets (Kalinić et al., 2021; Lee et al., 2020; Mohamed & Marzouk, 2021). Therefore, the main research questions of this studies are as follows: 1. Whether do user engagement and imagery affect in-person tour intentions? 2. What is the potential relationship between environment concern of touristic travel, in-person tour intentions and its possible influencing factors? 3. What is the linear and nonlinear models results of above relationships?

To address these gaps, our study aims at understanding the underlying impact of the systematic HCI user engagement drivers and imagery, and the EC influencing mechanism toward VR technologies on in-person tour intentions during COVID-19. Specifically, this study develops a theoretical framework based on the theory of user engagement integrating the Narrative transportation theory as well as EC perspective to explore the underlying antecedents and the influencing mechanism. This study is among the first, to our best knowledge, to assess the impact of the systematic human-computer interaction user engagement drivers (i.e., focused attention, perceived usability, aesthetic appeal, endurability, novelty, felt involvement), on individual VR technologies adopters and visitors' in-person tour intentions and the influencing mechanisms. Moreover, we examine crucial and yet not assessed mechanisms that the interaction effects of pro-environmental oriented EC in complementing, respectively, for the relationships of user engagement drivers and visitors' in-person tour intentions, and the relationships of imagery and visitors' in-person tour intentions. Then, a questionnaire-type survey was performed in China as a crucial emerging market, and the results were firstly tested with partial least squares (PLS) structural equation model (SEM). We considered both reflective and second-order formative constructs in PLS-SEM giving scope to their different advantages in a complex model since the user engagement drivers has six factors revealing the complexity. Meanwhile, the use of a hybrid PLS-SEM-artificial neural network (ANN) approach in assessing visitors' is a novel and hybrid approach in expert systems as in previous tourism studies. Moreover, previous PLS-SEM, SEM and ANN studies (Kalinić et al., 2021; Lee et al., 2020; Mohamed & Marzouk, 2021) cannot assess formative construct in the model. We proposed a novel and hybrid approach called Reflective and Formative PLS-SEM-ANN (FRPSA) with considering both reflective and second-order formative constructs in PLS-SEM giving scope to their different advantages in the complex model. The novelty of the FRPSA is among the first to integrate with a linear PLS model to predict VR technology usage and visitors' behavior in tourism in this study. This hybrid approach can overcome the deficiencies of the linear model while taking advantage of exploring nonlinear relationships. The novel approach has offered robustness against noise and improving the predictive power of the previously linear model. Subsequently, both theoretical and practical implications are discussed. The remainder is illustrated as follows. Section 2 demonstrates the theoretical framework and constructs our hypotheses. Section 3 depicts the methodology and assesses the results in Section 4. Section 5 discusses with the implications, conclusion, limitations, and future research directions. To illustrate the research progress vividly, we draft the flowchart of it in the Figure 1.

*Figure 1 here.*

## 2. Theoretical Background and Relationships Development

### *2.1 VR as the digital technology-enabled reality and its development in tourism*

Virtual reality (VR), as one of the essential technology-enabled reality realms, has a long history, though become popular recently with felt experience (Flavián et al., 2019). Several examples of the applications of these VR digital technologies can be found in tourism and related realm (Alyahya & McLean, 2021; Loureiro et al., 2020). According to emerging digital technologies, such a virtual world alters the means that operators and other stakeholders can understand their users or customers before, during, and after their experience (Kim et al., 2020). VR refers to a new means of machine interaction based on computer technology, modern simulation technology, and the latest sensor technology (Alyahya & McLean, 2021; Loureiro et al., 2020).

The core technology of VR is digital modeling, computer simulation, and human-computer interaction (Collange & Guegan, 2020). The fundamental characteristic of virtual reality is immersion-interaction-imagination, with the digital world of three-dimensional virtual space established by digital computer simulation, as well as the simulation of vision, hearing, touch, smell, balance sense, and other senses so that the experiencer can immerse himself in it, observe, feel and interact with the virtual three-dimensional space world in real-time and without restriction (Kim et al., 2020). The development history of VR technology is a typical true story of science fiction foresight inspiring scientific and technological innovation (Collange & Guegan, 2020). At the beginning of the 21st century, the one-chip computer has made significant progress in CPU processing, computing speed, capacity expansion, control parallel interface ability, etc.; force (force feedback) performance technology is becoming more and more mature; in 2012, oculus launched oculus rift, VR through crowdfunding; recently, Google, Samsung, and HTC released more device, and VR development ushered in a new round of peak (Huang et al., 2021). VR also appears more and more in tourism (Loureiro et al., 2020).

Immersive technologies such as virtual reality (VR) also enable travel operators and other related stakeholders to increase visitor satisfaction and engagement by providing unforgettable experiences (Barreda-Ángeles & Hartmann, 2022). The literature review shows that the relationship between the human-oriented consideration (e.g., ease of use and telepresence) and non-human technical aspects such as hardware and software has been conceptualized in different ways within the previous VR tourism studies (Huang et al., 2021; Loureiro et al., 2020; Reer et al., 2022). Moreover, the interactions between the human-oriented and the non-human-oriented technique (Kim et al., 2020) also were assessed by prior studies producing outcomes (e.g., VR tourism acceptance and Stimulus - Organize - Response process).

Further, the COVID-19 pandemic has a massive impact on all walks of life in countries around the world economy and added uncertainty (Fan et al., 2022; Lee et al., 2022; Lohr et al., 2022; Soulard & McGehee, 2022). The increasing number of people infected with the virus, the travel restrictions of tourists and the need for epidemic prevention and control, and the loss of income of people in many industries have a substantial negative impact on the tourism industry, which involves the movement of people (Kang et al., 2021; McLean & Barhorst, 2021). The revenue of tourist attractions and enterprises shrinks, and offline tourism entities are affected, but also online tourism enterprises are facing the same dilemma. In the past, many ideas of tourist attractions and enterprises are to pursue a large number of tourists and improve the comprehensive income of operation quantitatively. However, a large number of tourists tend to have a mediocre experience. After the epidemic, the quality of tourism products and services will improve accordingly. Based on the occasion of tourism, travel cannot be replaced. Attracting tourists through VR and other technological means has become the content that stakeholders need to explore. Some studies examine the visitors' social distancing perception between VR tour intention and real tour intention. The results show that social distancing affects visitors' virtual reality usage while negatively affecting in-person tours during the pandemic (Itani & Hollebeek, 2021; Wu et al., 2022).

Therefore, to summarize, VR research and commercial applications have flourished in academia and have led scholars to study the drivers of the success of technology-enabled reality in tourism. VR and related technologies are making a revolution in the way customers experience products and services. Yet, in related literature, no study of the current state of VR considers the impact of user engagement on tourists' behavior from the HCI perspective. Such immersive technologies have a core technology nature derived from human-computer interaction (Huang et al., 2021). VR is very the application of HCI, and the user engagement from the HCI perspective reflected by tourists and other individuals is the lack of assessment.

### 2.2 User engagement based on technology experience in human-computer interaction

In human-computer interaction and the philosophy of technology, the concept and framework of technology experience and user experience technology involvement are generated (Huang et al., 2021). The earliest theory of technological experience was put forward in the monograph Technological Experience published by MIT Press (O'Brien & Toms, 2008). Experience of Technology provides a theoretical basis for a more straightforward analysis of human experience by providing a new perspective and new way to view technology as experience. Starting from the pragmatism of empirical philosophers, they discuss the interaction between people and technology from the aspects of aesthetic input, situational creativity, value center, and meaning creation (O'Brien & Toms, 2010). The theory of technological experience reveals that technology is deeply embedded in daily experience in aesthetic, ethical, and functional ways. As a stage expression, they integrate the aesthetic experience into understanding technical experience. They offer a new way of looking at the specialized experience in a theory of technological experience that embraces creativity, openness, and relationship (O'Brien et al., 2018). When there is situational creativity in action, when every moment has potential, there is room for surprise; the experience of technology is sometimes frustrating and sometimes satisfying; however, the theory of technological experience provides a new way of looking at technology, indicating that the integration of technology into experience means that there are open elements and unfinished mechanisms, and only when technology is regarded as participating in sensory experience can it be fully understood (O'Brien & Toms, 2008).

Then, the cue framework of the dimension matrix of technological experience, combined with the theory of user engagement (UE) with technology, was proposed (O'Brien & Toms, 2008). Then they defined User engagement as a quality of user experience characterized by the depth of an actor's investment when interacting with a digital system (O'Brien et al., 2018, p28). User engagement is not just lead to user satisfaction. It is the ability to sustain engagement that can result in positive outcomes in many digital environments in previous studies, such as e-health, e-learning, and others (O'Brien et al., 2018; O'Brien & Lebow, 2013). User engagement toward VR technology will also potentially affect tourists' behavior from the HCI perspective.

## 2.3 Relationship development and Framework

Technology-enabled usage behavior is associated with the theory of reasoned action (TRA) as the theoretical framework basis since this theory is a well-accepted framework in technology diffusion (Alyahya & McLean, 2021; Hsiao & Chen, 2018). In tourism and related field, tour intention usually refers to a tourist's personal view or idea of things and the personal subjective thinking that results from it (Alyahya & McLean, 2021). The TRA has been continuously examined and expanded, and many empirical studies in VR have provided evidence for explanation (Alyahya & McLean, 2021; Hsiao & Chen, 2018). AR/VR users employ technology-enabled reality and terminal applications as tools for a virtual experience.

The human-computer interaction (HCI) community and technology management realm have drawn attention to understanding user engagement with many computer-mediated applications (Huang et al., 2021). The abstract nature of UE exacerbates the dynamic nature of computer-mediated interactions (O'Brien & Lebow, 2013). The user engagement scale (UES) was built with dimensions of focused attention, perceived usability, aesthetic appeal, endurability, novelty, and felt involvement (O'Brien & Toms, 2010). Focused attention refers to the feeling absorbed in the interaction and losing track of time; perceived usability refers to affect experienced as a result of the interaction and the degree of control and effort expended; aesthetic appeal refers to the attractiveness and visual appeal of the interface; endurability refers to the overall success of the interaction and users' willingness to recommend an application to others or engage with it in future; novelty refers to curiosity and interest in the interactive task; felt involvement refers to the sense of being "drawn in" and having fun (O'Brien et al., 2018, p30).

Focused attention derives from the users' perception of the passage of time and their awareness of what is happening outside of their interaction with digital technology (Huang et al., 2021; O'Brien et al., 2018). This condition is related to the user's ability to become engrossed and lost in the VR experience. Perceived usability pertains to the emotions experienced by respondents while completing VR tasks (O'Brien et al., 2018; O 'Brien & Toms, 2008). Related to these emotions is whether they take full advantage of the challenges required to experience

VR and how clear their perception of VR navigation is. Perceived usability also measures whether users feel they can perform the tasks they want to through VR and their perceived control over the interaction (O'Brien & Toms, 2010). The aesthetic appeal relates to specific features of the interface, such as screen layout and graphics/graphics, and respondents' overall aesthetic impression of VR appeal and sensory appeal (O'Brien et al., 2018). It is related to the visual appearance of the interface (Shang & Wu, 2019).

Endurability assesses respondents' likelihood of recommending the technology to others, as well as their likelihood of perceiving the experience as successful, rewarding, valuable, and on schedule (Huang et al., 2021; O'Brien and Lebow, 2013). Overall, this likelihood measured respondents' willingness to resume using the technology and recommend it to others, as well as their overall rating of the experience. Novelty assessed curiosity caused by participants' interest in using technical tasks. Arousing respondents' curiosity shows that, at different points in time, the experience contains surprising, unexpected, or new information. Felt involvement is a need-based cognitive or belief state, that is, psychological identification with an object based on an individual's significant needs and the perception that the object will satisfy those needs. This kind of involvement is adopted based on the user's perception of the importance of the experience, the level of importance or relevance to the pleasure of engagement and participation (O'Brien & Toms, 2008).

In-person tour intentions are defined as the motivations in the individual perception of the conscious decision process to perform a real offline tour (Alyahya & McLean, 2021). In VR tourism, consumers' perception and cognition of VR experience offer an experienced channel, which may influence the intent (Jeong et al., 2020). In a technology-supported real-world environment, most users adopt an experience-before-decision process based on the real-world environment (Kalinić et al., 2021). Therefore, if VR visitors feel focused on and felt involvement in what is happening in their VR interaction with digital technology, their positive beliefs will accumulate. At the same time, VR visitors' emotions are activated by the interviewees when they complete the task, acceptable perceived usability and endurability experiences are integrated with the psychological process, and their positive beliefs occur (Skard et al., 2021). In addition, if the screen layout and graphical/image interface, the virtual content of the travel site seems to become an aesthetic impression and novelty, attractive and sensory appeal will lead to sound beliefs about VR travel sites. Overall, these user engagement components from the human-computer interaction technology experience will affect visitors' stable psychological tendency towards a tourism VR to take an in-person tour. Thus, the following hypothesis is proposed:

H1. User engagement will positively influence in-person tour intentions.

Imagery is defined as thinking derived from quasi-conscious or perceptual experiences from which people reconstruct sensory information in their own minds and generate new ideas (Shang & Wu, 2019). This thought-generating process involves both unconscious responses and attention-forming processes under greater voluntary control. Previously, the distinction between perception and mental intentionality was ambiguous because they could often yield functionally comparable results. As a component of human behavior, the psychological intention has received more attention in psychology and learning research. Narrative transfer theory (Skard et al., 2021) formally uncovers the underlying mechanisms of psychological imagery. Narrative transportation is the concept that a "traveler" can experience psychologically returning to the journey (Shang & Wu, 2019). Previous research has shown that narratives enable recipients to share transformational experiences and have strong and lasting effects on psychological intentions (Shang & Wu, 2019; Skard et al., 2021). In travel VR conditions, users can create ongoing thoughts and ideas based on the UE experience; in the end, they can become fully "engaged" by being immersed in a specific travel VR environment. Therefore, travelers enter a quasi-perceptual state after experiencing VR through human-computer interaction, which can encourage travelers to share an almost real journey. This is the process of fitting similar realistic environments and mental activity states when people experience VR experience. This imagery process usually occurs automatically and unconsciously after the human-computer interaction UE experience.

In VR tourism, visitors' perception and cognition of the VR experience offer an imagery channel, which may influence the subsequent in-person tour intentions (Jeong et al., 2020; Skard et al., 2021). The imagery may influence individuals' intention to adopt offline tour patterns after conscious consideration of particular circumstances. More quasi consciousness accumulated from employing VR toward user engagement from HCI, and more active imagery and intention may occur (Skard et al., 2021). In other words, visitors' imagery could influence in-person tour intentions through a subjectively individual probability assessment based on the HCI engagement experience. The mediating effect of engagement-based imagery could predict the in-person tour intentions. Thus, the following hypotheses are proposed:

H2. Imagery will positively mediate the relationships between user engagement antecedents and in-person tour intentions, respectively.

The adverse environmental impacts of travel and tourism have not gone unnoticed (Hofman et al., 2022; Gössling, 2020). There were many studies on tourists, residents, and sustainable development, and the trend of research has gradually increased. Research explicitly highlights the role of travel in increasing pollution through carbon emissions (Talwar et al., 2022). In the past, there were many studies on tourists, local residents, and sustainable development, and the trend of research has gradually increased (Talwar et al., 2022; Suárez-Rojas et al., 2021). For example, through in-depth analysis and retrieval of literature, it can be found that the environmental impact of tourism waste is a relatively representative research category (Gössling, 2020; Hofman et al., 2022). The ongoing debate about the environmental effects of consumption has raised public awareness of the negative ecological consequences of consumption. As a consumption area, travelers are paying more and more attention to environment-friendly consumption and environmental protection behavior. Individuals' high awareness of the ecological impact of travel motivates them to think in a pro-environment way.

With the continuous development of the social economy, people's quality of life is also improving, and information technology and tourism integration are also developing (Suárez-Rojas et al., 2021; Talwar et al., 2022). Through VR experience, consumers can experience several different tourist destinations. Cloud tourism through VR (virtual reality) and other technologies will show the characteristics of scenic spots in the scenic area. The future of this new tourism technology will become a unique growth point for the tourism industry (Flavián et al., 2019; Loureiro et al., 2020). Various experience activities can also use AR, VR, and other technologies to provide exclusive props and costumes for tourists to enhance the reality of the experience (Bec et al., 2021; Zheng et al., 2021). The environment concern of touristic travel (EC) refers to people's anxiety and worry about the waste, emissions, and negative effects of global climate change that their tourism generates (Hedlund, 2011; Higgins-Desbiolles, 2020). It includes people's concern about environmental pollution and negative consequences for the sustainable development of the tourism they are engaged in (Higgins-Desbiolles, 2020; Suárez-Rojas et al., 2021). EC will provide a psychological space for environment-friendly tourists (Talwar et al., 2022). In this space, tourists compare different VR attractions through UE and often choose the few projects they most like to travel to in person. This means they have a choice of places they want to visit. In other words, EC enhances the transformation from UE to IPI. Consequently, an EC will positively moderate the relationships between UE and tourist attractions.

Similarly, with the development and application of tourism VR technology, tourism VR has been widely used to promote tourism products and services by scenic spots and enterprises (Kim et al., 2020; McLean & Barhorst, 2021). Suppose tourists experience the quasi-perception of psychological intention from VR information technology services. In that case, they can explore and seek information about scenic spots to find whether to enhance their willingness to make decisions on travel intentions. In the experience of different VR attractions, visitors tend to be younger and more affluent in age due to their active motivation and curiosity. These individuals may tend to experience mental intentions based on their characteristics and resources (Jeong et al., 2017). People often hope to pay a limited price to achieve environmental protection and ethical behavior (Shang & Wu, 2022; Suárez-Rojas et al., 2021). If individuals have a strong sense of moral orientation, they are more likely to engage in environmental actions and collective behavior. EC provides a unique individual moral context and brings environmental adjustment for ethical consumers' evaluation of tourist attractions (Talwar et al., 2022). Therefore, if VR can almost achieve the site's entire experience, the eco-tourist can satisfy his curiosity and leisure through imaginative or creative thinking. Thus, EC may neutralize the imagery, and this interaction mechanism will weaken the willingness of the ITI. Consequently, an EC will negatively moderate the relationships between imagery and tourist attraction. Based on these statements, we propose the following hypotheses, and the research framework and rationale are shown in Figure 2.

H3a. Environment concern of touristic travel will positively moderate the relationships between VR and in-person tour intentions.

H3b. Environment concern of touristic travel will negatively moderate the relationships between imagery and in-person tour intentions.

*Figure 2 here.*

## 3. Methodology

We adopted a two-stage method to carry out the research, including the first stage of empirical research and the second stage of simulation experiment based on a neural network of machine learning. In the first stage of empirical research, a questionnaire-based survey was conducted to collect the first stage empirical data. There are eight variables as constructs derived from the related literature (Hedlund, 2011; Higgins-Desbiolles, 2020; Itani & Hollebeek, 2021; Jeong et al., 2020; O'Brien et al., 2018; O'Brien & Toms, 2010; Shang & Wu, 2019). All items were measured with a 5-point Likert scale, ranging from "1 = strongly disagree" to "5 = strongly agree" in Survey adapted from the previous literature (Hedlund, 2011; Higgins-Desbiolles, 2020; Itani & Hollebeek, 2021; Jeong et al., 2020; O'Brien et al., 2018; O'Brien & Toms, 2010; Shang & Wu, 2019). The demographic profiles regarding age, education, and gender were listed at the end of the questionnaire. After the survey instrument was developed, it was examined by seven VR scholars. Considering the comments, the expressions were improved. We performed a pre-test with 43 experienced VR tourists in a tourist attraction in Harbin City, China, and the results of the pre-test manifested the acceptable criteria of scale to conduct the large-scale Survey.

This study is mainly aimed at releasing VR tourists. Participants are invited to fill in a questionnaire and will receive a survey allowance after completing the questionnaire. The questionnaires were collected face-to-face using a quota sampling method, with target demographic data matching the results of the most recent regional consumer census based on age, gender, and education level of adults 18 years and older (Hair et al., 2017; Shang & Wu, 2022). The questionnaire survey was conducted from August to December 2021, and a total of 1,000 questionnaires were distributed. Due to incomplete data, 91 questionnaires were excluded from the 706, and 615 were considered valid. This is to obtain an appropriate sample size to meet the data collection rules (Hair et al., 2017; Shang & Wu, 2022). Male and females accounted for 41.8% and 58.6%. The majority of respondents were 18-35 years old (59.8 %), followed by those over 36 years old (40.2%). More than half hold a bachelor's degree or below (88.7%), followed by a graduate degree or below (11.3%).

According to Hair et al., (2017), VBSEM method is commonly used for statistical analysis in both VR and tourism research, and has the advantage of overcoming multicollinearity (Chen and Lin, 2015; Hair et al., 2017). At the same time, since formative construction has the advantage of systematic overall evaluation, apart from regular reflective constructs, this paper contains user engagement as the second-order formative construct and the formative construct can only be tested by variance-based SEM. Thus, partial least squares structural equation model (PLS-SEM) as the variance-based SEM was used to evaluate the model.

The second phase of this study used simulation experiments by machine learning techniques. Because SEM can test the hypothesis of causality, it cannot detect nonlinear relationships. However, artificial neural networks are suitable for research environments with prediction. ANN is a flexible nonlinear analysis method, and it is inappropriate to test assumptions (Chong, 2013; Kalinić et al., 2021). The sample size in this paper is 50 times the number of adjustable parameters, which can be used as the minimum sample size for ANN (Leong et al., 2019). Artificial neural networks are network structures consisting of many artificially designed processing units that are abstracted and simulated by workers in the field of expert systems based on the basic features of the human brain (Chong, 2013; Kalinić et al., 2021). ANNs have been supervised learning models that study the characteristics of human thinking and simulate intelligent human behavior based on the physiological structure of the human brain. However, hypothesis assessment is impossible because of the "black box" operation of ANN. Therefore, this paper proposes two stages to take advantage of strength and overcome limitations. Consequently, we propose a method called "Reflective and Formative PLS-SEM-ANN (FRPSA)," a new methodological contribution to expert systems and AI research.

## 4. Results

### *4.1 Measurement Model*

Three steps measurement model criteria for Reflective and Formative PLS-SEM were utilized between potential structures and indicators. For the first step, the tests to check for convergence validity were conducted (see Table 1) (Chen & Lin, 2015; Shang and Wu, 2022). Cronbach's alpha is 0.7 or greater. Composite reliability was greater than 0.7. The mean variance of extraction was greater than 0.5. The results show that all indicators meet the requirements and represent the recognized reliability. According to the PLS discriminant validity analysis standard, we tested whether the factor load itself was higher than the cross load of other items in the same level. Table 2 shows the corresponding discriminant validity (Hair et al., 2017).

For the second step, using SmartPLS 3.0 programming to correct for the potential effects of common method variance (CMV), the complete collinearity test was performed (Chen & Lin, 2015; Hair et al., 2017). The results of the variance inflation factor, including outer loadings, are all less than 5 reflecting the fulfillment of the complete collinearity test standard, and the data have no measurement error and CMV in reflective and formative PLS-SEM measurement assessment (Hair et al., 2017; Shang and Wu, 2022). Moreover, for the third step, the significance test of the outer weights of the formative indicators the outer weights were assessed. The results in Table 3 show that the outer weights of all formative indicators are significant in statistical point estimation with P value lower than 0.001 (Hair et al., 2017). Confidence intervals bias-corrected test results also denote no bias in assessing statistical interval estimation. Thus, Table 3 shows an acceptable significance of the level of the formative indicator in this study.

*Table 1 here.*
*Table 2 here.*

*Table 3 here.*

### 4.2 Structural Model

Structural models are used to test relationships between various constructs and test hypotheses by assessing the significance of path coefficients (Chen & Lin, 2015). We resampled the bootstrapping technique through SmartPLS 3.0 programming and considered 5000 repetitions to obtain a stable estimate (Chen & Lin, 2015; Hair et al., 2017; Shang and Wu, 2022). Table 4 depicts the results of the differently structured models, showing the dependent variable only model (Model 1), the intermediate model (Model 2), and the control variable and comprehensive model (Model 3), used to analyze in-person tour intentions as the dependent variable's intention to visit in person.

*Table 4 here.*
*Table 5 here.*

All direct variables of the user engagement antecedents from the second-order formative construct in Model 1. Each of the antecedents, focused attention ($\beta = 0.171$, $p < 0.001$), perceived usability($\beta = 0.182$, $p < 0.001$), aesthetic appeal ($\beta = 0.181$, $p < 0.001$), endurability ($\beta = 0.185$, $p < 0.001$), novelty ($\beta = 0.178$, $p < 0.001$), and felt involvement ($\beta = 0.179$, $p < 0.001$) has the significantly positive influence on user engagement and results show that the second-order formative construct was well established. Meanwhile, the intermediate path assessment was partially put in Model 2, and the results show that user engagement has a significant relationship with imagery ($\beta = 0.916$, $p < 0.001$). Furthermore, we performed an analysis based on the bootstrapping method (Hair et al., 2017; Shang and Wu, 2019). A resampling test performed 5,000 repetitions to achieve stable estimates (Hair et al., 2017). Table 5 shows the results, and all relationships of effects are significant, indicating a mediating effect concerning imagery, thus supporting H2.

The control variables and comprehensive assessment were tested in Model 3. User engagement as the antecedent has a significantly positive influence on in-person tour intentions ($\beta$ = 0.800, p < 0.001), thus supporting H1. We also entered moderating terms in the estimation equation. We found that the interaction between the relationship of user engagement and EC positively impacts in-person tour intentions ($\beta$ = 0.128, p < 0.01), thereby supporting H3a. It was also found that the interaction between the relationship of imagery and EC has a negative impact on in-person tour intentions ($\beta$ = −0.095, p < 0.05), thereby supporting H3b. In other words, the EC fails to positively moderate the relationships of imagery and in-person tour intentions. Consistent with the literature, the demographic variables, including age, gender, education level, and perceived pandemic context, i.e., social distancing based on COVID-19 and perceived susceptibility based on COVID-19, were set as control variables in this study (Itan and Hollebeck, 2021; Leong et al., 2019). For the control variables assessment, we also found that social distancing based on COVID-19 and perceived susceptibility based on COVID-19 were significant for affecting in-person tour intentions thus setting control variables in the context of the COVID-19 pandemic could be necessary for this and related studies.

### 4.3 Neural network experiment results

The research on the neural network has appeared very early, and the neural network has been a prominent and multi-disciplinary field (Mohamed & Marzouk, 2021). Early scientists of neuroscience and mathematics were inspired by what they knew about biological neural networks: that when each neuron connected to other neurons "fired," it sent chemicals to the connected neurons that changed the electrical potential in those neurons; If a neuron becomes more positive than a "threshold," it activates, "fires up," sending chemicals to other neurons. Multilayer perceptron (MLP), as suggested by artificial neural networks, are used in this paper (Kalinić et al., 2021). The disadvantage of the single-layer perceptron is that it can only solve the problem of linear separable classification. To increase the classification ability of the network, the only way is to use a multilayer network, adding a hidden layer between the input and output layers to form a multilayer perceptron (Kalinić et al., 2021; Mohamed & Marzouk, 2021). In this paper, a multilayer perceptron is a kind of multilayer feedforward network model composed of three parts: perceptron elements consisting of an input layer, a hidden layer of one or more computes nodes, and an output layer of compute nodes.

Similar to Leong et al., (2020), the significant paths from the PLS testing, focused attention, perceived usability, aesthetic appeal, endurability, novelty, felt involvement, user engagement, imagery, and environment concern as input neurons were set into the neural network analysis. The number of neurons in the input layer is the dimension of the input signal, which is 9 in this paper. The number of hidden layers and hidden nodes can be set. Based on previous technology and related combined prediction model research and the hidden layer number calculation rules, input layer neurons are brought into the integer part function (Kalinić et al., 2021). Finally, the number of hidden layers is determined as one layer. In this paper, the in-person tour intention is the last spirit element in the output layer. The excitation function of each neuron in a multilayer perceptron is an activation function sigmoid (Kalinić et al., 2021; Lee et al., 2020), as shown below:

$$f(x) = \frac{1}{1 + e^{-x}}$$

(1)

The principle is that neural networks introduce an activation function. Each layer is equivalent to Taylor's expansion of the input model of the previous layer to obtain higher-order characteristics based on the activation function. In fact, for the input model of the last layer, this is also a regression process, but the independent variables are changed. For the whole input model, higher-order approximations are obtained. On the whole, its essence is a regression process. As the maximum entropy principle is the cornerstone of probabilistic learning, the neural network needs to approximate the system's entropy to the maximum point. The activation function plays a gradual, cumulative role. Thus, a continuous prediction model is fitted. Multilayer feed-forward neural networks can approach continuous functions of arbitrary complexity with arbitrary precision only with enough hidden layers of neurons.

This study uses a tenfold cross-validation routine to avoid overfitting (Kalinić et al., 2021) to obtain the root mean square error (RMSE) from training and testing processes based on sigmoid activation functions. Subsequently, after evaluation of the hidden and output layers activation functions, a tenfold cross-validation routine is used to avoid overfitting (Lee et al., 2020) to obtain the square sum error (SSE) and RMSE from training and testing evaluation. Based on the adopted ANN model. Since the mean value of the RMSE is relatively low (Table 6), it can be concluded the ultimate model offers good and acceptable predictions (Chong, 2013; Mohamed & Marzouk, 2021).

The percentage of variance explained by the ANNs is assessed (Kalinić et al., 2021; Lee et al., 2020). R squared is the variance of the expected output. The results showed that input neurons predicted 94.62% of the variance of in-person tour intentions. The proposed PLS-ANN model explains 94.62% of in-person tour intentions as the variance output variable. To determine the relative influence and importance of each predictive variable, sensitivity analysis was conducted by ANN. Based on these values of analysis, the normalized importance of each predictive variable can be evaluated among all predictive variables (Kalinić et al., 2021; Lee et al., 2020). The normalized importance (NI) of all predictive variables as the input neurons was based on sensitivity analysis divided by the average importance indicated in Table 7. It found that felt involvement was the strongest predictor of in-person tour intentions, followed with aesthetic appeal (NI = 92.1%), perceived usability (NI = 77.6%), imagery (NI = 75.0%), user engagement (NI = 74.4%), focused attention (NI = 61.9%), novelty (NI = 58.0%), endurability (NI = 51.0%), and environment concern (NI = 44.4%).

## 5. Discussion and Conclusion

### *5.1 Theoretical Contribution*

This paper presents several essential contributions to academic research from the theoretical perspective. First, this study discusses the influencing factors of tourists' VR experience and in-person tour intentions and proposes an integrated theoretical framework of systematic HCI user engagement and in-person tour intentions. Most research on VR is oriented toward technological perception, individual psychological traits, or interaction context (Bec et al., 2021; Merkx & Nawijn, 2021; Skard et al., 2021). However, VR as an information technology HCI is its core technology. It is scarce to analyze VR from the perspective of systematic HCI, especially in tourism. HCI user engagement has explored online shopping, online learning, and other fields (Jeong et al., 2020; O'Brien et al., 2018), but the features of purchasing products, learning scenarios, and tourism are pretty different. The visitor experience is often associated with leisure and entertainment and curiosity-oriented expectations. Previous studies on online shopping, online learning, and other fields do not apply directly to tourists, while the research on HCI user engagement exploring tourists' VR attitudes and in-person tour intentions in tourism is lacking. This research establishes a research framework based on empirical data testing. The results show that user engagement elements, namely felt involvement, aesthetic appeal, perceived usability, focused attention, endurability, and novelty, directly affect in-person tour intentions. Tourists' VR imagery is an intermediary between in-person tour intentions and user engagement. Therefore, this paper identifies the influence mechanism of user engagement on tourists' VR imagery and in-person tour intentions to make specific theoretical contributions to relevant knowledge. This has enriched the research literature on HCI user engagement and tourism and improved the understanding of the indirect effects of user engagement, visitor VR imagery, and intention on each other. Therefore, the HCI user engagement and narrative transportation theory were also expanded and enriched.

Second, this study identifies the influence mechanism of the EC factor on tourists' VR imagery, user engagement, and in-person tour intentions. It identifies the asymmetric relationships in the moderating effects. Previous studies in tourism and environment studies show that EC may seriously affect users' experiences and feelings (Jeong et al., 2017; Shang & Wu, 2019). However, in the context of tourism VR, we have not yet been clear whether EC will affect tourists' VR imagery, user engagement, and in-person tour intentions. We investigate interaction effects. The interaction between the relationship of user engagement and EC positively impacts in-person tour intentions. It was also found that the interaction between the relationship of imagery and EC has a negative impact on in-person tour intentions. The EC fails to positively moderate the relationships of imagery and in-person tour intentions. Travelers who pay attention to environmental protection pay more attention to UE experience but pay less attention to narrative transmission quasi perceptual state. In other words, the experience of VR is just right, and there is no need to pursue the trigger and maintenance of quasi perceptual state, which is quite different from some domain characteristics, such as online learning. Therefore, this study focuses on the systematic HCI user engagement of tourists and EC and concludes that users will perceive different experience levels and quasi perceptions. Thus, our research paper helps fill in some gaps in the research and identifies the EC as the significant moderator of user engagement and in-person tour intentions in visitor experience.

In addition, this study discusses the application of PLS-ANN analysis. It provides a new research method for the academic community, namely, tourist intention and HCI user participation in virtual tourism reality, which can capture linear and nonlinear relationships in several ways. Existing VR and tourism VR research mainly applied linear models (e.g., PLS-SEM, SEM, linear regression) (Itan & Hollebeek, 2021; Jeong et al., 2020). Meanwhile, previous PLS-SEM, SEM, and ANN studies (Kalinić et al., 2021; Lee et al., 2020; Mohamed & Marzouk, 2021) cannot assess formative construct in the model. Although some studies have used PLS-ANN (Leong et al., 2019), to our best knowledge, the difference is that we proposed a novel and hybrid method called Reflective and Formative PLS-SEM-ANN (FRPSA) with considering both reflective and second-order formative constructs in PLS-SEM giving scope to their different advantages in a complex model. This is among the first studies to propose and test this new FRPSA approach. Compared with the formative constructs, the traditional covariance-based SEM estimation reflective constructs pointer model is not applicable since model settings will have problems. Evaluating second-order formative constructs can make the relationship between path and structure clearer and more focused. Thus, after taking advantage of reflective and second-order formative constructs in PLS-SEM assessment, combining these techniques with ANN can further improve prediction accuracy. Through evaluating linear and nonlinear models, the felt involvement and aesthetic appeal show both the linear significance impact and nonlinear importance. Therefore, researchers should attach importance to improving and promoting felt involvement and aesthetic appeal from a systematic HCI user engagement perspective. Thus, it reveals valuable information about researchers and management action with evaluation and prediction by balancing linear significance and nonlinear importance evaluation for user engagement. The new approach can serve as a guide for other research environments. It can address the complexity and oversimplification of visitor decision-making processes and problems associated with nonlinear models.

## 5.2 Practical Implications

The research results provide a practical reference for tourism VR activities. First, given economic development and social progress, more people have realized the importance of sustainable development. Customers' awareness of green perception has gradually emerged and improved. Suppose tourism enterprises and scenic spot operations focus on environmental protection and sustainable development in their operational processes and implement appropriate publicity for products and services that are sustainably valuable. In that case, visitors will prefer these products and services. For example, some tourists are worried about the carbon emissions brought by transportation during travel. Tourism enterprises can provide tourists with more green transportation tools, such as electric vehicles, shared bicycles, etc. This can reduce the environmental and ethical guilt brought by tourists. Furthermore, when enterprises or organizations donate or implement an appropriate budget to causes that align with their development strategies, they demonstrate a better commitment to corporate social responsibility and environment strategy. A better reputation and brand image of enterprises and other organizations would therefore be established among target tourist groups (e.g., ethical consumers) because of this type of marketing and operation strategy.

### 5.3 Conclusions

The contribution of this study to tourism VR and the diffusion of VR technology lies in that it puts forward an integrated theoretical framework, adopts the method of combining statistics and machine learning, and explores the influence brought by the environment concern of touristic travel, imagery, and HCI user engagement by combining tourists' psychological perception and experience. To improve the efficiency and effectiveness of destination marketing and tourism management under the impact of COVID-19, this study expands and extends the content level covered by systematic HCI user engagement and narrative transportation theory in the tourism VR context. This paper strengthens the theoretical basis of VR technology and VR tourism in the following ways. Firstly, it identifies the elements of HCI user engagement, including felt involvement, aesthetic appeal, perceived usability, focused attention, endurability, and novelty, which all affect tourists' attitudes toward VR. HCI user engagement is explored for the first time in a systematic way as an important prerequisite for VR technology and VR tourism. Secondly, tourists' imagery toward VR plays an intermediary role in the relationship between HCI user engagement and in-person tour intentions. It helps to improve the understanding of the indirect effects of HCI user engagement, imagery, and in-person tour intentions on each other. Thirdly, the study identifies the asymmetric influence mechanism of environment concern factor on tourists' VR attitudes and in-person tour intentions. The interaction between the relationship of user engagement and environment concern of touristic travel has a positive impact on in-person tour intentions. Finally, we proposed a novel and hybrid method called Reflective and Formative PLS-SEM-ANN (FRPSA) with testing, which can capture linear and nonlinear relationships and further improve the accuracy of prediction. The felt involvement and aesthetic appeal show both the linear significance impact and nonlinear importance. Thus, it reveals useful information about researchers and management action with evaluation and prediction by balancing linear significance and nonlinear importance evaluation for analyzing decision-making processes.

In addition, based on the following reasons, this paper also has some limitations. First, the samples studied are targeted at tourists in China's emerging markets, but different countries will have different cultures and lifestyles. Second, sampling techniques can be further optimized, which may lead to a certain degree of bias. Further studies will consider collecting longitudinal data to enhance the robustness of conclusions. Third, the influencing factors are also one of the limiting factors. This research paper proposes and explores HCI user engagement and domain segmentation innovativeness factors. Other latent variables related to epidemic perception factors during epidemics can also be considered in subsequent studies. To facilitate the development of better strategies for the tourism VR experience, further research may uncover potential differences in age, gender, and educational background among different groups. Future research should follow this approach to improve visitors' attitudes, intentions, and behaviors in VR situations.

**Tables**

**Table 1. Reliability Test Results of PLS-SEM**

|  | Cronbach's Alpha | Composite Reliability | Average Variance Extracted |
|---|---|---|---|
|  | (1) | (2) | (3) |
| AE | 0.827 | 0.920 | 0.852 |
| EC | 0.931 | 0.956 | 0.878 |
| EN | 0.890 | 0.948 | 0.901 |
| FA | 0.881 | 0.944 | 0.893 |
| FI | 0.886 | 0.946 | 0.898 |
| IMG | 0.865 | 0.917 | 0.787 |
| ITI | 0.922 | 0.950 | 0.865 |
| NO | 0.935 | 0.968 | 0.939 |
| PU | 0.900 | 0.953 | 0.909 |
| UE | 0.973 | 0.976 | 0.775 |

Note: FA: focused attention; PU: perceived usability; AE: aesthetic appeal; EN: endurability; NO: novelty; FI: felt involvement; UE: user engagement; IMG: imagery; EC: environment concern of touristic travel; ITI : in-person tour intentions.

**Table 2. Discriminant Validity**

| | UE | IMG | EC | ITI |
|---|---|---|---|---|
| | (1) | (2) | (3) | (4) |
| FA1 | 0.948 | 0.833 | 0.867 | 0.848 |
| FA2 | 0.942 | 0.857 | 0.860 | 0.792 |
| PU1 | 0.956 | 0.824 | 0.819 | 0.828 |
| PU2 | 0.952 | 0.778 | 0.765 | 0.799 |
| AE1 | 0.918 | 0.744 | 0.766 | 0.784 |
| AE2 | 0.928 | 0.824 | 0.785 | 0.840 |
| EN1 | 0.947 | 0.793 | 0.797 | 0.794 |
| EN2 | 0.952 | 0.831 | 0.855 | 0.878 |
| NO1 | 0.968 | 0.752 | 0.753 | 0.797 |
| NO2 | 0.970 | 0.775 | 0.787 | 0.807 |
| FI1 | 0.949 | 0.853 | 0.846 | 0.880 |
| FI2 | 0.946 | 0.801 | 0.779 | 0.838 |
| IMG1 | 0.733 | 0.869 | 0.735 | 0.688 |
| IMG2 | 0.794 | 0.898 | 0.811 | 0.856 |
| IMG3 | 0.846 | 0.893 | 0.830 | 0.778 |
| EC1 | 0.852 | 0.830 | 0.940 | 0.812 |
| EC2 | 0.872 | 0.834 | 0.943 | 0.820 |
| EC3 | 0.844 | 0.853 | 0.929 | 0.808 |
| ITI1 | 0.773 | 0.783 | 0.754 | 0.919 |
| ITI2 | 0.827 | 0.830 | 0.830 | 0.934 |
| ITI3 | 0.820 | 0.833 | 0.834 | 0.936 |

Note: FA: focused attention; PU: perceived usability; AE: aesthetic appeal; EN: endurability; NO: novelty; FI: felt involvement; UE: user engagement; IMG: imagery; EC: environment concern of touristic travel; ITI : in-person tour intentions.

**Table 3.** The significance test of the outer weights of the formative indicators

| | Original Sample | T Statistics | P Values | Confidence intervals Bias | Confidence intervals 2.5% | Confidence intervals 97.5% |
|---|---|---|---|---|---|---|
| | (1) | (2) | (3) | (4) | (5) | (6) |
| FA1<-UE | 0.097 | 81.533 | 0.000 | 0.000 | 0.095 | 0.099 |
| FA2<-UE | 0.094 | 72.704 | 0.000 | 0.000 | 0.091 | 0.096 |
| PU1<-UE | 0.096 | 84.673 | 0.000 | 0.000 | 0.094 | 0.098 |
| PU2<-UE | 0.092 | 73.392 | 0.000 | 0.000 | 0.089 | 0.094 |
| AE1<-UE | 0.089 | 62.614 | 0.000 | 0.000 | 0.086 | 0.092 |
| AE2<-UE | 0.096 | 82.219 | 0.000 | 0.000 | 0.094 | 0.099 |
| EN1<-UE | 0.093 | 80.074 | 0.000 | 0.000 | 0.091 | 0.095 |
| EN2<-UE | 0.099 | 79.522 | 0.000 | 0.000 | 0.097 | 0.102 |
| NO1<-UE | 0.091 | 53.772 | 0.000 | 0.000 | 0.087 | 0.094 |
| NO2<-UE | 0.093 | 55.978 | 0.000 | 0.000 | 0.089 | 0.096 |
| FI1<-UE | 0.095 | 77.061 | 0.000 | 0.000 | 0.093 | 0.098 |
| FI2<-UE | 0.100 | 81.244 | 0.000 | 0.000 | 0.098 | 0.102 |

Note: FA: focused attention; PU: perceived usability; AE: aesthetic appeal; EN: endurability; NO: novelty; FI: felt involvement; UE: user engagement; IMG: imagery; EC: environment concern of touristic travel; ITI : in-person tour intentions.

**Table 4. Structural Model**

| | β estimate (T statistics) | | |
| --- | --- | --- | --- |
| | UE | IMG | ITI |
| | (1) | (2) | (3) |
| **Control variables** | | | |
| Age | | | 0.007* |
| | | | (0.446) |
| Education | | | 0.003 |
| | | | (0.177) |
| Gender | | | 0.028 |
| | | | (1.408) |
| Social distancing based on COVID-19 | | | 0.575*** |
| | | | (12.102) |
| Perceived susceptibility based on COVID-19 | | | 0.334*** |
| | | | (6.714) |
| Independent variables | | | |
| FA | 0.171*** | | |
| | (78.419) | | |
| PU | 0.182*** | | |
| | (89.356) | | |
| AE | 0.181*** | | |
| | (128.709) | | |
| EN | 0.185*** | | |
| | (81.316) | | |
| NO | 0.178*** | | |
| | (56.647) | | |
| FI | 0.179*** | | |
| | (87.508) | | |
| UE | | 0.916*** | 0.800*** |
| | | (128.709) | (13.830) |
| IMG | | | 0.117** |
| | | | (2.672) |
| **Interaction Terms** | | | |
| EC× UE | | | 0.128** |
| | | | (2.846) |
| EC× IMG | | | -0.095* |
| | | | (2.307) |

Key: *p-value < 0.05, **p-value < 0.01, ***p-value < 0.001.

Note: FA: focused attention; PU: perceived usability; AE: aesthetic appeal; EN:

endurability; NO: novelty; FI: felt involvement; UE: user engagement; IMG: imagery; EC: environment concern of touristic travel; ITI : in-person tour intentions.

**Table 5. Indirect Test Results of PLS-SEM**

| Relationship | Path coefficient | T-Values | P-Values | Supported |
|---|---|---|---|---|
| | (1) | (2) | (3) | (4) |
| UE -> IMG -> ITI | 0.107 | 2.669 | 0.008 | YES |
| AE -> UE -> IMG -> ITI | 0.018 | 2.669 | 0.008 | - |
| EN -> UE -> IMG -> ITI | 0.019 | 2.666 | 0.008 | - |
| FA -> UE -> IMG -> ITI | 0.019 | 2.659 | 0.008 | - |
| FI -> UE -> IMG -> ITI | 0.020 | 2.668 | 0.008 | - |
| NO -> UE -> IMG -> ITI | 0.019 | 2.683 | 0.007 | - |
| PU -> UE -> IMG -> ITI | 0.019 | 2.660 | 0.008 | - |

Key: *p-value < 0.05, **p-value < 0.01, ***p-value < 0.001.

Note: FA: focused attention; PU: perceived usability; AE: aesthetic appeal; EN: endurability; NO: novelty; FI: felt involvement; UE: user engagement; IMG: imagery; EC: environment concern of touristic travel; ITI : in-person tour intentions.

**Table 6 Test of SSE and RMSE**

| | Training | | | Testing | | | Training Time |
|---|---|---|---|---|---|---|---|
| | N | SSE | RMSE | N | SSE | RMSE | (Second) |
| | (1) | (2) | (3) | (4) | (5) | (6) | (7) |
| NO.1 | 544 | 4.953 | 0.167 | 71 | 0.502 | 0.134 | 0.190 |
| NO.2 | 546 | 5.394 | 0.180 | 69 | 0.786 | 0.230 | 0.170 |
| NO.3 | 552 | 4.907 | 0.164 | 63 | 0.431 | 0.124 | 0.180 |
| NO.4 | 555 | 4.793 | 0.156 | 60 | 0.530 | 0.219 | 0.180 |
| NO.5 | 553 | 4.228 | 0.140 | 62 | 0.646 | 0.197 | 0.180 |
| NO.6 | 551 | 3.584 | 0.120 | 64 | 0.528 | 0.148 | 0.240 |
| NO.7 | 547 | 3.973 | 0.131 | 68 | 0.579 | 0.187 | 0.160 |
| NO.8 | 553 | 4.252 | 0.141 | 62 | 0.591 | 0.174 | 0.190 |
| NO.9 | 543 | 5.012 | 0.167 | 72 | 0.623 | 0.193 | 0.160 |
| NO.10 | 556 | 4.183 | 0.138 | 59 | 0.431 | 0.137 | 0.230 |
| Average | 550 | 4.528 | 0.150 | 65 | 0.502 | 0.134 | 0.188 |
| St. dev. | 4 | 0.535 | 0.018 | 4 | 0.786 | 0.230 | 0.026 |

Note: FA: focused attention; PU: perceived usability; AE: aesthetic appeal; EN: endurability; NO: novelty; FI: felt involvement; UE: user engagement; IMG: imagery; EC: environment concern of touristic travel; ITI : in-person tour intentions.

**Table 7 Sensitivity Analysis**

| Table s | Tabl es | Table s | Table s | Tabl es | Table s | Table s | Table s | Table s |
|---------|---------|---------|---------|---------|---------|---------|---------|---------|
| Table s | Tabl es | Table s | Table s | Tabl es | Table s | Table s | Table s | Table s |
| Table s | Tabl es | Table s | Table s | Tabl es | Table s | Table s | Table s | Table s |
| Table s | Tabl es | Table s | Table s | Tabl es | Table s | Table s | Table s | Table s |
| Table s | Tabl es | Table s | Table s | Tabl es | Table s | Table s | Table s | Table s |
| Table s | Tabl es | Table s | Table s | Tabl es | Table s | Table s | Table s | Table s |
| Table s | Tabl es | Table s | Table s | Tabl es | Table s | Table s | Table s | Table s |
| Table s | Tabl es | Table s | Table s | Tabl es | Table s | Table s | Table s | Table s |
| Table s | Tabl es | Table s | Table s | Tabl es | Table s | Table s | Table s | Table s |
| Table | Tabl | Table | Table | Tabl | Table | Table | Table | Table |

|  | FA | PU | AE | EN | NO | FI | UE | IMG | EC |
|---|---|---|---|---|---|---|---|---|---|
|  | (1) | (2) | (3) | (4) | (5) | (6) | (7) | (8) | (9) |
| NO.1 | 0.098 | 0.122 | 0.145 | 0.080 | 0.091 | 0.158 | 0.117 | 0.118 | 0.070 |
| NO.2 | 0.118 | 0.081 | 0.097 | 0.115 | 0.107 | 0.078 | 0.130 | 0.140 | 0.135 |
| NO.3 | 0.146 | 0.096 | 0.123 | 0.082 | 0.098 | 0.110 | 0.110 | 0.113 | 0.123 |
| NO.4 | 0.126 | 0.075 | 0.151 | 0.120 | 0.082 | 0.152 | 0.113 | 0.101 | 0.080 |
| NO.5 | 0.135 | 0.056 | 0.151 | 0.068 | 0.160 | 0.141 | 0.122 | 0.101 | 0.066 |
| NO.6 | 0.110 | 0.098 | 0.149 | 0.043 | 0.124 | 0.198 | 0.070 | 0.115 | 0.093 |
| NO.7 | 0.112 | 0.070 | 0.133 | 0.056 | 0.142 | 0.178 | 0.105 | 0.104 | 0.099 |
| NO.8 | 0.099 | 0.059 | 0.148 | 0.085 | 0.137 | 0.156 | 0.095 | 0.112 | 0.110 |
| NO.9 | 0.152 | 0.089 | 0.129 | 0.113 | 0.100 | 0.105 | 0.099 | 0.127 | 0.088 |
| NO.10 | 0.139 | 0.086 | 0.139 | 0.059 | 0.135 | 0.184 | 0.070 | 0.092 | 0.095 |
| Average importance | 0.124 | 0.083 | 0.137 | 0.082 | 0.118 | 0.146 | 0.103 | 0.112 | 0.096 |
| Normalized importance | 61.90 % | 77.60 % | 92.10 % | 51.00 % | 58.00 % | 100.00 % | 74.40 % | 75.00 % | 44.40 % |

Note: FA: focused attention; PU: perceived usability; AE: aesthetic appeal; EN: endurability; NO: novelty; FI: felt involvement; UE: user engagement; IMG: imagery; EC: environment concern of touristic travel; ITI : in-person tour intentions.